\documentclass[openacc]{rstransa}

\newtheorem{ass}{\bf Assumption}[section]

\newcommand{\bX}{\mathbf{X}}

\newcommand{\bY}{\mathbf{Y}}

\newcommand{\bZ}{\mathbf{Z}}
\newcommand{\bW}{\mathbf{W}}

\newcommand{\barZ}{\bar{Z}}
\newcommand{\barz}{\bar{z}}

\newcommand{\bU}{\mathbf{U}}

\newcommand{\one}{\mathbf{1}}

\newcommand{\mate}{\tau^\textsc{m}}
\newcommand{\pate}{\tau^\textsc{p}}
\newcommand{\sate}{\tau^\textsc{s}}
\newcommand{\cate}{\tau(x)}
\newcommand\ci{\perp\!\!\!\perp}

\def\ci{\perp\!\!\!\perp}
\def\mis{\textup{mis}}
\def\obs{\textup{obs}}
\def\d{\textup{d}}
\def\co{\textup{co}}
\def\nt{\textup{nt}}
\def\at{\textup{at}}
\def\df{\textup{df}}

\def\sumN{\sum_{i=1}^N}

\def\ipw{\textup{ipw}}
\def\reg{\textup{reg}}
\def\dr{\textup{dr}}
\def\hattauipw{\widehat{\tau}^{\ipw}}
\def\hattaureg{\widehat{\tau}^{\reg}}
\def\hattaudr{\widehat{\tau}^{\dr}}

\DeclareMathOperator\bE{\mathbb E} 

\begin{document}

\title{Bayesian Causal Inference: A Critical Review}

\author{
Fan Li$^{1}$, Peng Ding$^{2}$, and Fabrizia Mealli$^{3}$}

\address{$^{1}$Duke University, Durham, NC, USA\\
$^{2}$University of California, Berkeley, CA, USA\\
$^{3}$University of Florence and EUI, Florence, Italy
}

\subject{machine learning, statistics}


\corres{Fan Li\\
\email{fl35@duke.edu}}

\begin{abstract}
This paper provides a critical review of the Bayesian perspective of causal inference based on the potential outcomes framework. We review the causal estimands, assignment mechanism, the general structure of Bayesian inference of causal effects, and sensitivity analysis. We highlight issues that are unique to Bayesian causal inference, including the role of the propensity score, the definition of identifiability, the choice of priors in both low and high dimensional regimes. We point out the central role of covariate overlap and more generally the design stage in Bayesian causal inference. We extend the discussion to two complex assignment mechanisms: instrumental variable and time-varying treatments. We identify the strengths and weaknesses of the Bayesian approach to causal inference. Throughout, we illustrate the key concepts via examples. 
\end{abstract}


\begin{fmtext}
\section{Introduction} \label{sec::intro}
Causality has long been central to the human philosophical debate and scientific pursuit. There are many relevant questions, e.g. the philosophical meaning of causation or deducing the causes of a given phenomenon. Among these questions, statistics---which concerns measurements---arguably can contribute the most to the question of measuring the effects of causes. Statistics infers associations between variables. Even though the research questions in many statistics-based studies are causal in nature, a first lesson in elementary statistics is that \emph{association does not imply causation}. Distinguishing between causation and spurious association between various events is a challenging task in science. Broadly speaking, statistical causal inference is about building a framework that \emph{(i)} defines causal effects under general scenarios, \emph{(ii)} specifies assumptions under which one can identify causation from association, and \emph{(iii)} assesses the sensitivity to the causal assumptions and finds ways to mitigate.

\end{fmtext}
\maketitle

A mainstream statistical framework for causal inference is the potential outcomes framework \cite{Neyman23, Rubin74}. Under this framework,  following the dictum ``\emph{no causation without manipulation}'' \cite{Rubin75}, a \emph{cause} is a pre-specified \emph{treatment} or \emph{intervention} that is at least hypothetically manipulable. A typical causal question is  ``would an individual have a better outcome had he took treatment $A$ versus treatment $B$?'' Causal effects are defined as comparisons of potential outcomes, also known as counterfactuals, under different treatment conditions for the same units. The main hurdle to interpreting the association between the treatment and the outcome as a causal effect is confounding, i.e., the presence of factors that are associated with both the treatment and the outcome. For example, patients with worse health conditions may be more likely to obtain a beneficial medical treatment; then directly comparing the outcomes of the treated and control patients, without adjusting for the difference in their baseline health conditions, would bias the causal comparisons and mistakenly conclude that the treatment is harmful. Randomized  experiments, known as \emph{A/B testing} in industry or randomized controlled trials in medicine, are the gold standard for casual inference by eliminating confounding via randomization. But modern causal inference has increasingly relied on  observational data. The potential outcomes framework provides basis for identifying and estimating causal effects---quantities defined based on counterfactuals---from the factual data in the presence of confounding, using randomized or observational data. This framework is applicable to a wide range of problems in many disciplines and has been increasingly adopted in the era of machine learning. Other frameworks for causal inference, including the causal diagram \cite{PearlBook} and invariant prediction \cite{peters2016causal}, are beyond the scope of this review.

There are three primary inferential approaches within the potential outcomes framework \cite{ding2018causal}: Fisherian randomization test, Neymanian repeated-sampling evaluation, and Bayesian inference. The first two approaches belong to the Frequentist paradigm and have  been dominant, with many popular tools such as propensity scores, matching, and weighting. The Bayesian approach has several established advantages for general statistical analysis, including automatic uncertainty quantification, coherently incorporating prior knowledge, and offering a rich collection of advanced models for complex data. As causal studies increasingly involve real-world big data, there has been a recent surge of research in Bayesian inference of causal effects \cite{hill2011bayesian, ZiglerDominici14, hahn2018regularization,hahn2020bayesian,linero2022how}, but there lacks a comprehensive appraisal of the current state of the research. This paper aims to fill this gap. Due to the space limit, we do not intend to provide a catalog of the existing research on this topic, but rather discuss the big picture of why and how to conduct Bayesian causal inference in general settings. We emphasize the unique questions, challenges and opportunities that the Bayesian approach brings to causal inference. We hope this review can stimulate broader and deeper cross-fertilization between causal inference and Bayesian analysis. 

Section \ref{sec::primitives} introduces the preliminaries of the potential outcomes framework, and briefly discusses several Frequentist methods to causal inference. Section \ref{sec::BayesianCI_general} outlines the general structure of Bayesian causal inference, focusing on ignorable treatment assignments at one time point. Section \ref{sec::outcomemodel} discusses model specification and implications in high dimensional settings. Section \ref{sec::PSrole} reviews the role and various uses of the propensity score in Bayesian causal inference. Section \ref{sec::SA} outlines sensitivity analysis in observational studies. Section \ref{sec::extensions} describes two complex assignment mechanisms: instrumental variable and time-varying treatments. Section \ref{sec::discussion} concludes.

\section{Estimands, Identification, and Frequentist Estimation} \label{sec::primitives}

To convey the main ideas, we focus on the case with a binary treatment at one time period, which can be readily extended to multiple treatments and multiple time points.
Consider a sample of units drawn from a target population, indexed by $i\in \{1,...,N\}$. Each unit can potentially be assigned to one of two treatment levels $z$, with $z=1$ for the active treatment and $z=0$ for the control. Let $Z_i (=z)$ be the binary variable indicating unit $i$'s observed treatment status. For unit $i$, a vector of $p$ covariates $X_i$ are observed before the treatment, and an outcome $Y_i$ is observed after the treatment. A confounder is a pre-treatment variable that is associated with both the treatment and the outcome; it can be observed, as a subset of the covariates $X_i$, or unobserved. Below we use covariates and confounders interchangeably. We use the $A \ci B \mid C$ notation to denote conditional independence between two variables $A$ and $B$ given variable $C$\cite{dawid1979conditional}. We also use the bold font to indicate a vector consisting of the corresponding variables for the $N$ units, e.g. $\bZ=(Z_1,\ldots,Z_N)'$.  

We maintain the standard stable unit treatment value assumption (SUTVA) \cite{Rubin80}, namely, there is \emph{(i)} no different version of a treatment, and \emph{(ii)} no interference in the sense that one unit's potential outcomes are not affected by other units' treatment assignment. Under SUTVA, each unit $i$ has two potential outcomes:  $Y_i(1)$ and $Y_i(0)$. Causal effects are contrasts of potential outcomes under different treatment conditions for the same set of units. The individual treatment effect (ITE) for unit $i$ is $\tau_i=Y_i(1) - Y_i(0)$. Averaging $\tau_i$ over a sample we obtain the sample average treatment effect (SATE): 
$\tau^\textsc{s} \equiv N^{-1} \sumN\tau_i$. 
Furthermore, the conditional average treatment effect (CATE) is the average of the individual treatment effect of all units with the covariate value $x$:
\begin{equation}
    \tau(x)\equiv\bE\{ Y_i(1)-Y_i(0)\mid X_i=x\} =\mu_1(x)-\mu_0(x), \label{def:CATE}
\end{equation}
where $\mu_z(x)\equiv \bE\{ Y_i(z) \mid  X_i=x \}$ for $z=0,1$. 
Averaging $\tau_i$ or $\tau(X_i)$ over a target population gives the population average treatment effect (PATE): 
\begin{equation}
\tau^\textsc{p}  \equiv  \bE\{ Y_i(1)-Y_i(0)\} \label{def:PATE}
= \bE\{\tau(X_i)\} .
\end{equation}
The PATE is a function of the distribution of the potential outcomes in a population, whereas the SATE is a function of the potential outcomes themselves. The subtle distinction in their definitions leads to important differences in inferential and computational strategies, as will be discussed later. Traditionally, the SATE is of interest in randomized experiments where the target population is the specific sample, whereas the PATE is of interest in observational studies where the target population is the population from which the sample is drawn. In general, the choice of a causal estimand is determined by the scientific question in hand rather than statistical considerations. Note that although both the ITE and  CATE are important in characterizing treatment effect heterogeneity, they are obviously different; however, these two estimands are sometimes conflated in the literature.   

The \emph{fundamental problem of causal inference} \cite{holland1986causal} is that, for each unit only the potential outcome corresponding to the actual treatment, $Y_i^\obs \equiv Y_i=Y_i(Z_i)$, is observed or \emph{factual}, and the other potential outcome, $Y_i^\mis=Y_i(1-Z_i)$, is missing or \emph{counterfactual}. Therefore, additional assumptions are necessary to identify  the causal effects. The key identifying assumptions concern the assignment mechanism, i.e. the process that determines which units get what treatment and hence which potential outcomes are observed or missing \cite{Rubin78}. The vast majority of causal studies assume certain versions of an \emph{ignorable assignment mechanism}, where the treatment assignment is independent of the potential outcomes conditional on some observed variables. Specifically, in the simple setting of a binary treatment at one time, ignorability consists of two sub-assumptions \cite{Rubin78, rosenbaum1983central}. 

\begin{ass}\label{ass::ignorability} (\textbf{Ignorability}). 
\textbf{(a) Unconfoundedness}. 
$\Pr\{  Z_i \mid  Y_i(0), Y_i(1), X_i \}  = \Pr( Z_i \mid  X_i )$, or equivalently $Z_i\ci \{Y_i(0), Y_i(1)\} \mid X_i$. \textbf{(b) Overlap}. $0< e(X_i) <1 $ for all $i$, where  $e(x)\equiv \Pr(Z_i=1\mid X_i=x)$ is the propensity score \cite{rosenbaum1983central}. 
\end{ass}
The unconfoundedness assumption states that there is no unmeasured confounding, and the overlap assumption states that each unit has non-zero probability of being assigned to each treatment condition. These two assumptions together ensure that the conditional distribution of the potential outcomes is identifiable from observed data as 
\begin{equation}
\mu_z(x) \equiv  \bE\{ Y_i(z) \mid  X_i=x  \} = \bE( Y_i \mid  Z_i = z,  X_i=x), \quad \mbox{for all }  z, x. \label{eq:identification}
\end{equation}
Therefore, the CATE is identified as $\tau(x)=\mu_1(x)-\mu_0(x)$, and the PATE is identified as $\tau^\textsc{p}  = \bE\{\mu_1(X_i)-\mu_0(X_i)\}$. This underlines the estimation strategy of \emph{outcome modelling}: we can specify a model  for the outcome function $\mu_z(x)$, and estimate the CATE by $\widehat{\tau}(x)=\widehat{\mu}_1(x)-\widehat{\mu}_0(x)$, and the PATE by 
$
\hattaureg = N^{-1} \sumN \{ \widehat \mu_1(X_i) -  \widehat \mu_0(X_i) \} ,
$
where $\widehat\mu_z(x)$  is the estimated outcome model from the observed data. 

In randomized experiments, the treatment assignment is known and controlled by the experimenters, and ignorability holds by design. In observational studies, the treatment assignment is unknown and uncontrolled, and ignorability at best holds approximately. A key concept in causal inference is \emph{overlap and balance}, which refers to the similarity in the distributions of the covariates between the comparison groups. In general, as the two groups become more balanced, the causal estimates become less sensitive to the estimate strategy and model specification. In the ideal case of a randomized experiment, all---measured and unmeasured---covariates are balanced in expectation, i.e., they have the same multivariate distribution in the two treatment arms. Consequently, the simple difference-in-means estimator, $\widehat{\tau}=\{\sumN Y_iZ_i/\sumN Z_i\}- \{\sumN Y_i(1-Z_i)/\sumN (1-Z_i)\}$, is unbiased for $\tau^\textsc{s}$ and $\tau^\textsc{p}$; furthermore, even a misspecified linear outcome model leads to a consistent estimate of $\tau^\textsc{s}$ \cite{lin2013agnostic}. On the contrary, in observational studies, the two groups are often imbalanced in many covariates, e.g. patients receiving the treatment may be generally sicker and older than those receiving the control. In such cases, directly comparing the difference in the outcomes between the two groups would give biased estimates of the causal effects. Moreover, the fit of an outcome model would rely on extrapolation in the regions where the two groups are poorly overlapped, and consequently outcome-model-based estimators, such as $\hattaureg$, are sensitive to the model specification. Therefore, a main effort in causal inference with observational data is to ensure overlap and balance to mimic a randomized experiment as closely as possible. This process does not involve the outcome and is referred to as the \emph{design} stage, in contrast to the \emph{analysis} stage, which utilizes the outcome and estimates causal effects given the design stage \cite{Rubin07}. A causal analysis of an observational study usually has both design and analysis stages, in parallel with those of a randomized controlled experiment. 

The propensity score plays a central role in causal inference with observational data, owing to its two special properties \cite{rosenbaum1983central}. First, the propensity score is a balancing score in the sense that $Z_i \ci e(X_i) \mid X_i$. This means that balancing the scalar propensity score balances the multivariate distribution of the covariates. Second, if a treatment assignment is unconfounded given $X_i$, then it is unconfounded given $e(X_i)$, that is,  $ Z_i \ci \{ Y_i(0), Y_i(1)\}\mid X_i$ implies $ Z_i \ci \{ Y_i(0), Y_i(1)\}\mid e(X_i)$. In observational studies, $e(X_i)$ is usually unknown and needs to be estimated, e.g., via a logistic regression model of the treatment on the covariates.

The propensity score is usually used with matching, weighting, or stratification to achieve balance and estimate causal effects. Specifically, matching methods use a certain algorithm to find pairs of units in the two groups with similar covariates according to a distance metric, e.g. the propensity score or the Mahalanobis distance, and then calculate the difference in the average observed outcome between the groups in the matched sample \cite{rubin2006matched, abadie2006matching, AbadieImbens11}. Weighting methods assign a weight to each unit, so that the weighted distribution of the covariates in the two groups are balanced \cite{li2018balancing}, and then calculate the weighted difference in the outcomes between the two groups. An important weighting scheme is inverse probability weighting (IPW), based on the identification formula of the PATE:
$  \tau^\textsc{p}  =  \bE\{  Z_iY_i / e(X_i)- (1-Z_i)Y_i / (1-e(X_i)) \}.$
A corresponding IPW estimator \cite{rosenbaum1987model} is 
$
\hattauipw = N^{-1} \sumN  \{ Z_iY_i /  \widehat{e}(X_i)  
-  (1-Z_i) Y_i / ( 1-\widehat{e}(X_i) ) \},
$ 
where $\widehat{e}(X_i)$ denotes the estimated propensity score for unit $i$. One can further augment the IPW estimator by an outcome model to obtain a semiparametric efficient estimator \cite{robins1994estimation}:
$
\hattaudr = \hattaureg  +  N^{-1} \sumN \{  Z_i R_i / \widehat{e}(X_i) -   (1-Z_i) R_i / ( 1-\widehat{e}(X_i) )   \}, 
$
where $R_i = Y_i - \widehat \mu_{Z_i}(X_i)$ is the residual from the outcome model. The IPW estimator $\hattauipw $ is consistent for $\tau$ if the propensity score model is correct, and the outcome-model estimator $\hattaureg$ is consistent if the outcome model is correct. Because the bias of the estimator $\hattaudr$ is a product of the residual of the propensity score model and that of the outcome model,  $\hattaudr$ is  \emph{doubly robust} in the sense that it is consistent if either the propensity score or the outcome model, but not necessarily both, is correctly specified \cite{Bang05}. Despite the seemingly different construction, matching estimators, with proper mathematical formulations, can be viewed as nonparametric versions of $\hattauipw$, $\hattaureg $ and $\hattaudr$ based on nearest-neighbor regressions \cite{lin2021estimation}. These are the main Frequentist estimation strategies for $\tau^\textsc{p}$ under ignorability.  When the target estimand is the CATE, the primary estimation strategy is outcome modelling. We will discuss how to specify the outcome model in Section \ref{sec::outcomemodel}\ref{sec::specification}.

\section{General Structure of Bayesian Causal Inference} 
\label{sec::BayesianCI_general}
\subsection{Basic Factorization and Versions of Causal Estimands}
Because of the unavoidable missing potential outcomes, causal inference under the potential outcomes framework is inherently a missing data problem \cite{Rubin78, ding2018causal}. The Bayesian paradigm offers a unified framework for statistical inference with missing data and thus for causal inference \cite{Rubin76}. Below we review the general structure of Bayesian causal inference that was first outlined in \cite{Rubin78}. 
 
Four quantities are associated with each unit $i$, $\{Y_i(0), Y_i(1), Z_i, X_i\}$, where $\{Z_i, X_i,Y_i(Z_i)\}$ are observed but $Y_i(1-Z_i)$ is missing. Bayesian inference views all these quantities as random variables and centers around specifying a model for them. Based on the Bayesian model, we can draw inference on causal estimands---functions of the model parameters, covariates and potential outcomes---from the posterior predictive distributions of the parameters and the unobserved potential outcomes. Specifically, we assume the joint distribution of these random variables of all units is governed by a parameter $\theta = (\theta_X, \theta_Z, \theta_Y)$, conditional on which the random variables for each unit are \emph{i.i.d.}. Then we can factorize the joint density $\Pr\{ Y_i(0),Y_i(1),Z_i,X_i \mid \theta\} $ for each unit $i$ as
\begin{eqnarray}
\Pr\{ Z_i \mid Y_i(0),Y_i(1), X_i; \theta_{Z}\} 
 \cdot    \Pr\{ Y_i(0),Y_i(1)\mid X_i; \theta_{Y}\}
 \cdot    \Pr(X_i ; \theta_X).  \label{eq::bayesian-joint}
\end{eqnarray}
The three terms in \eqref{eq::bayesian-joint} represent the model for the assignment mechanism, potential outcomes, and covariates, respectively. Under ignorability, the assignment mechanism further reduces to the propensity score model $\Pr(Z_i \mid X_i; \theta_{Z})$. 

Before diving into the technical details, we first clarify the subtle but important difference between the Bayesian estimation of the PATE and SATE estimands. For the PATE, we rewrite the outcome-model-based identification formula in Section \ref{sec::primitives} as $\tau^\textsc{p} = \int  \{ \mu_1(x; \theta_Y) - \mu_0(x; \theta_Y) \} F(\text{d}x; \theta_X)$, which depends only on the unknown parameters $\theta_X$ and $\theta_Y$. Therefore, Bayesian inference for the PATE requires obtaining posterior distributions of $(\theta_X, \theta_Y)$. In contrast, the SATE $\tau^\textsc{s}$ is a function of the potential outcomes $\{Y_i(0), Y_i(1)\}_{i=1}^N$, which involves both observed and missing quantities. Bayesian inference for the SATE requires imputing the missing potential outcomes $Y_i^{\mis}$ from their posterior predictive distributions based on the outcome model, and consequently deriving the posterior distribution of  $\tau^\textsc{s}.$ 

However, in practice, we rarely model the possibly multi-dimensional covariates $X_i$, and instead condition on the observed values of the covariates. This is equivalent to replacing $F(x; \theta_X)$ with $ \widehat{\mathbb{F}}_X$, the empirical distribution of the covariates. Therefore, most Bayesian causal inference (for example \cite{chib2007analysis,hill2011bayesian}) in fact focuses on the mixed average treatment effect (MATE) \cite{ding2018causal}: 
\begin{eqnarray} \label{est:con_ace}
\tau^\textsc{m} \equiv \int  \tau(x; \theta_{Y} ) \widehat{\mathbb{F}}_X(\d x)
= N^{-1} \sum_{i=1}^N   \tau  (X_i; \theta_{Y}) ,
\end{eqnarray}
where $\tau(x; \theta_{Y} )=\tau(x)$ highlights the dependence on the parameter $\theta_{Y}.$ The MATE is a convenient approximation of the PATE and is particularly natural under the Bayesian paradigm. The difference between the MATE and SATE is subtle: the former equals the average of the CATE whereas the latter equals the average of the ITEs over the finite sample. Based on the posterior distributions, the PATE has the largest uncertainty, whereas the SATE has the smallest uncertainty. The distinction between these estimands is illustrated in the following example. 

\textbf{Example 1} (Covariate adjustment in a randomized experiment) Consider a completely randomized experiment with covariates $X$. Assume the true model for  potential outcomes is
$$
\begin{pmatrix}
Y_i(1)\\
Y_i(0)
\end{pmatrix} \mid (X_i, \beta_1,\beta_0,\sigma_1^2,\sigma_0^2, \rho )
\sim  \mathcal N\left(
\begin{pmatrix}
\beta_1' X_i\\
\beta_0' X_i
\end{pmatrix},
\begin{pmatrix}
\sigma_1^2 & \rho \sigma_1 \sigma_0 \\
\rho \sigma_1 \sigma_0 & \sigma_0^2
\end{pmatrix}
\right) ,\quad i= 1,\ldots, N.
$$
This model implies two univariate normal marginal models: $Y_i(z)\mid X_i, \beta_z, \sigma_z^2 \sim \mathcal N(\beta_z' X_i, \sigma_z^2)$ for $z=0, 1$. In this example, the CATE is $\tau(x) = (\beta_1 - \beta_0)' x$; the PATE, SATE and MATE are
\begin{equation}
\tau^\textsc{p}=  (\beta_1-\beta_0)'\bE(X_i),\quad
\tau^\textsc{s}=  N^{-1}\sum_{i=1}^N\{ Y_i(1)-Y_i(0) \},\quad
\tau^\textsc{m} =(\beta_1-\beta_0)'\bar{X}, \label{eq:ATE-closedform}
\end{equation}
respectively, where $\bar{X} = N^{-1}\sum_{i=1}^N X_i$ is the sample mean of the covariates.  $\Box$

\subsection{Posterior Inference of Causal Effects}
Regardless of the version of the target estimand, the following assumption is commonly adopted.

\begin{ass}\label{ass::prior-independence}
\textbf{(Prior independence)}. The parameters for the models of assignment mechanism $\theta_Z$, outcome $\theta_Y$, and covariates $\theta_X$ are a priori distinct and independent. 
\end{ass} 

Assumption \ref{ass::prior-independence} imposes independent prior distributions for parameters $(\theta_X, \theta_Z, \theta_Y)$. It is unique to the Bayesian paradigm of causal inference. It is imposed primarily for modelling and computational convenience and may appear innocuous. However, as elaborated in Section \ref{sec::outcomemodel}\ref{sec::high-dim}, it may lead to unintended and undesirable implications in high-dimensional problems. Under Assumptions \ref{ass::ignorability} and \ref{ass::prior-independence}, the joint posterior distribution of $ \theta = (\theta_X, \theta_Z, \theta_Y)$ and the missing potential outcomes is proportional to
\begin{eqnarray}
\Pr(\theta_X)  \prod_{i=1}^N \Pr(X_i; \theta_X)
 \cdot    \Pr(\theta_Z) \prod_{i=1}^N  \Pr(Z_i \mid X_i; \theta_Z)
 \cdot   \Pr(\theta_Y)  \prod_{i=1}^N  \Pr\{ Y_i(1), Y_i(0)\mid X_i; \theta_Y\} . \label{eq::posterior_theta}
\end{eqnarray}
From \eqref{eq::posterior_theta}, the posterior distributions of $\theta_X$ and $\theta_Y$, and consequently of $\pate$, do not depend on the second component corresponding to the propensity score. Therefore, the propensity score model is \emph{ignorable} in Bayesian inference of $\pate$. The same ignorability argument applies to other estimands such as $\sate$, $\mate$ and $\cate$ \cite{Rubin78, hill2011bayesian, ding2018causal}. Furthermore, inference of $\tau^\textsc{m}$ does not depend on the covariate model $\Pr(X_i; \theta_X)$. Because of this, it is essential to specify the outcome model $\Pr\{ Y_i(1), Y_i(0)\mid X_i; \theta_Y\} $ in Bayesian causal inference.

By definition, $\pate=\bE\{ Y_i(1)\}-\bE\{Y_i(0)\}$  does not depend on the association between $Y_i(0)$ and $Y_i(1)$, denoted by the parameter $\rho$. Similarly, $\cate$ does not depend on $\rho$, but $\sate$ does. So in the inference of $\pate$ and $\cate$, we usually directly specify the marginal models $\Pr\{ Y_i(z) \mid X_i; \theta_Y \}$ or equivalently $\Pr(Y_i \mid Z_i=z, X_i; \theta_Y)$ under ignorability \cite{chib2007analysis}. The observed-data likelihood based on \eqref{eq::posterior_theta} becomes $\prod_{i: Z_i = 1}  \Pr(Y_i\mid Z_i=1, X_i; \theta_Y) \prod_{i: Z_i = 0}  \Pr(Y_i\mid Z_i=0, X_i; \theta_Y).$ Imposing a prior for $\theta_Y$, we can proceed to infer $\theta_Y$ and subsequently $\pate$, $\mate$, or $\cate$ using the usual Bayesian inferential procedures. 

Bayesian inference of $\sate$ is more complex, because it depends on both $Y_i(0)$ and $Y_i(1)$ and thus requires posterior sampling of both $\theta_Y$ and $\bY^{\mis}$. The most common sampling strategy is through data augmentation: iteratively simulate  $\theta$ and $\bY^\mis$ given each other and the observed data, namely from $\Pr(\theta_Y \mid\bY^{\mis}, \bY^{\obs}, \bZ, \bX)$ and $\Pr(\bY^{\mis}\mid \bY^{\obs}, \bZ, \bX; \theta_Y)$. The former, given the observed data and the imputed $\bY^\mis$, can be straightforwardly obtained by a complete-data analysis based on $\Pr\{  \theta_{Y} \mid   \bY(1), \bY(0) , \bX \} \propto \Pr(\theta_{Y}) \prod_{i=1}^N  \Pr\{ Y_i(1), Y_i(0) \mid  X_i;  \theta_{Y} \}$. The latter requires more elaboration. Specifically, we can show that $\Pr(\bY^{\mis}\mid \bY^{\obs}, \bZ, \bX; \theta_Y)$ is proportional to $\prod_{i: Z_i=1}   \Pr\{    Y_i(0) \mid  Y_i(1),  X_i; \theta_{Y}  \}  \prod_{i: Z_i=0}    \Pr\{   Y_i(1) \mid  Y_i(0),  X_i;  \theta_{Y} \}.$ This renders that imputing the $\bY^\mis$ depends crucially on the joint distribution of $\{ Y_i(1), Y_i(0) \}$. Because $Y_i(0)$ and $Y_i(1)$ are never jointly observed, the data provide no information about their association $\rho$. Unless the specific marginal model places constraints on $\rho$, the posterior distribution of $\rho$ would be the same as its prior. Consequently, the posterior distribution of $\sate$ would be sensitive to the prior of $\rho$.  

The above discussion prompts us to clarify the notion of identifiability in Bayesian inference. Under the Frequentist paradigm, a parameter is identifiable if any of its two distinct values  give two different distributions of the observed data. Under the Bayesian paradigm, there is no consensus. For example, Lindley \cite{lindley1972bayesian} argued that all parameters are identifiable in Bayesian analysis because with proper prior distributions, posterior distributions are always proper. In this sense, $\rho$ is identifiable. However, due to the fundamental problem of causal inference, there is no information in the data on $\rho$ and it is reasonable to label it as nonidentifiable. This is distinct from the parameters that the data provide direct information on, e.g. those in the marginal distributions of the outcomes in each arm, which are reasonable to label as identifiable. Lindley's perspective of identifiability blurs such distinction. A more informative perspective is provided by Gustafson \cite{gustafson2015bayesian}, who argued that a parameter is \emph{weakly} or \emph{partially} identifiable, if a substantial region of its posterior distribution is flat, or its posterior distribution depends crucially on its prior distribution even with large samples, such as $\rho$. Another example of a partially identifiable parameter is $\Pr\{Y_i(1)>Y_i(0)\}$, which depends on $\rho$ \cite{lu2018treatment, ding2018causal}. In this perspective, identifiability in Bayesian inference is no longer all-or-nothing; instead it is a continuum in between. This issue motivates the strategy of \emph{transparent parameterization}, where one separates identifiable and non-identifiable parameters, and treat the latter as sensitivity parameters in a sensitivity analysis \cite{daniels2008missing, gustafson2009limits, richardson2010transparent, ding2016potential, franks2019flexible}. More discussion will be given in Section \ref{sec::SA}\ref{sec::SA-outcome}.

\textbf{Example 1 revisited} We now illustrate the posterior inference of the causal estimands in Example 1. Here the parameters $\beta$'s and $\sigma$'s are identifiable, but $\rho$ is not in the Frequentist sense. We fit a Bayesian linear regression model of $Y_i$ on $X_i$ to each observed arm $z$, with independent priors on $( \beta_1, \sigma_1^2)$ and $(\beta_0, \sigma_0^2)$. The observed likelihood factorizes into two parts: the data in treatment group $\{  (X_i, Y_i) : Z_i = 1 \}$ and control group $\{ (X_i, Y_i) : Z_i = 0 \}$ contribute to the likelihood of $(\beta_1, \sigma_1^2)$ and  $(\beta_0, \sigma_0^2)$, respectively.  For example, imposing the conventional conjugate normal-inverse $\chi^2$ priors, we can draw from the posterior distribution of $\beta$ and $\sigma$, and thus that of the MATE by plugging the posterior draws into the closed-form of $\tau^\textsc{m}$ in \eqref{eq:ATE-closedform}.  To obtain the PATE, we would have to specify a multivariate model for $\Pr(X; \theta)$, and derive the posterior distribution of $\theta_X$ and then plug it into the closed form of $\tau^\textsc{p}$ in \eqref{eq:ATE-closedform}. This can also be implemented, e.g. via a Bayesian bootstrap step without a model, as described in the next paragraph. To obtain the SATE, we could specify a prior for $\rho$ or fix it to a value. Given $\rho$ and each draw of $( \beta_1, \beta_0, \sigma_1^2, \sigma_0^2)$, we can impute $Y_i^{\mis}$ as follows: for treated units, $Y_i^{\mis}=Y_i(0)$, and we draw
$
Y_i(0)
$ from $\mathcal  N(   \beta_0' X_i + \rho  \sigma_0 / \sigma_1 \cdot  (  Y_i - \beta_1' X_i ) , \sigma_0^2(1-\rho^2)   )
$; for control units, $Y_i^{\mis}=Y_i(1)$, and we draw
$
Y_i(1) $ from $ \mathcal  N(   \beta_1' X_i + \rho \sigma_1 / \sigma_0 \cdot  (  Y_i - \beta_0' X_i ) , \sigma_1^2(1-\rho^2)   ).
$
Plugging these posterior predictive draws of $Y_i^{\mis}$ and the observed outcomes into the definition of $\tau^\textsc{s}$, we obtain its posterior distribution. We suggest varying the sensitivity parameter $\rho$ from $0$ to $1$, which corresponds to conditionally independent potential outcomes and perfectly correlated potential outcomes, respectively. 
$\Box$

An interesting alternative Bayesian strategy is through the Bayesian bootstrap \cite{Rubin81}, where the units are re-weighted with weights drawn from a Dirichlet distribution. The Bayesian bootstrap is a general strategy to simulate the posterior distribution of a parameter under a nonparametric model, which can be viewed as the limit of the inference under the Dirichlet Process prior \cite{ferguson1973}. This renders the Bayesian bootstrap to be relevant to causal inference in at least two ways. First, one can generate posterior samples from the distribution of $\Pr(X_i;\theta_X)$ without specifying a parametric model. This is desirable in inferring the population estimands like the PATE and the CATE \cite{oganisian2020hierarchical}. However, how to integrate these samples of $X$ into the inference of the target causal estimand is case-dependent and generally adds complexity to the analysis compared to the MATE. Second, the Bayesian bootstrap offers a general recipe for incorporating many standard Frequentist procedures into Bayesian inference. For example, Taddy et al. \cite{taddy2016nonparametric} used it to quantify the uncertainty in linear and tree-based methods for estimating the CATE. Chamberlain and Imbens \cite{Chamberlain2003} used it in M-estimation with an application to the setting of instrumental variables (see Section \ref{sec::extensions}\ref{sec::principalstratification}). However, we view the Bayesian bootstrap approach as peripheral to Bayesian causal inference because it does not capitalize on arguably the main strength of Bayesian inference, namely, a unified inferential framework underpinned by the Bayes theorem with versatile choice of priors and outcome models.

\section{Model Specification} \label{sec::outcomemodel}

\subsection{Common Specification of the Outcome Model} \label{sec::specification}
Section \ref{sec::BayesianCI_general} shows that the central component in Bayesian inference of the CATE, PATE, and MATE is to specify the outcome model $\mu_z(x)=\bE (  Y_i \mid X_i=x, Z_i=z; \theta_Y ) $. We can either model the two treatment groups jointly with a single function $\mu_z(x) = \mu(z, x)$ or model each group separately with two functions $\mu_1(x)$ and $\mu_0(x)$, known as S-learner or T-learner, respectively, in the literature \cite{kunzel2019metalearners}. The most basic outcome model is a linear regression: $\mu(z, x)=  x+z+xz$, with Gaussian error terms, where the treatment-covariate interaction term  $xz$ captures the treatment effect heterogeneity. This model is equivalent to specifying a linear regression in each group. But the equivalence does not hold for nonlinear models; in fact, S-learners and T-learners with the same type of nonlinear models often lead to markedly different causal estimates. 

Linear regressions are easy to implement and interpret, but they are often too restricted. In real world problems, it is crucial to specify $\mu( z, x)$ flexibly enough to approximate the possibly complex underlying true data generating mechanism. This is particularly desirable as the recent focus in causal inference has been moving toward heterogeneous treatment effects. Outcome modelling is the most natural approach in these studies: one can simply specify an outcome model and derive the CATE as a function of the model parameters. There has been a rapidly increasing adoption of nonparametric and machine learning models for $\mu( z, x)$. One of the most widely used such models bases on regression trees. At a high level, regression trees partition the covariate space into non-overlapping regions and the prediction in each region is based solely on the data that fall in that region. The parameters of a regression tree characterize where to split the covariate space and how to predict the outcomes in a terminal node \cite{breiman2017cart}. An ensemble of regression trees---usually referred to as forests---are often combined to improve the prediction. 

Within the Bayesian paradigm, the Bayesian Additive Regression Tree (BART) \cite{chipman2010bart} has become very popular for causal inference. BART places certain priors on the parameters of the regression trees to control the depth of the tree and the degree of shrinkage of the mean parameters in terminal nodes. Hill \cite{hill2011bayesian} first advocated to use BART to specify the outcome model $\mu(z, x)$ in a S-learner. One can also specify a T-learner with a separate BART model for each treatment group $\mu_z(x)$. However, without any additional structure on the marginal models $\mu_z(x)$, T-learners often result in large variance of the treatment effects. Hahn et al. \cite{hahn2020bayesian} proposed the \emph{Bayesian Causal Forest} method based on an alternative reparametrization,  $\mu(z, x)=g_1(x)+g_2(x)z$, where $g_1(x)$ models the distribution of $Y(0)$ and $g_2(x)$ represents the heterogeneous treatment effect, with a separate BART prior for $g_1(x)$ and $g_2(x)$. The BART models have a number of advantages, including fast computation, good performance of  default choice of hyperparameters and available software. When a study has adequate covariate overlap, BART has been shown to outperform numerous competing methods, including (the Frequentist) random forests, in many empirical applications, e.g. \cite{dorie2019automated,hu2021estimating}. Other Bayesian nonparametric models, such as Gaussian process \cite{ray2020semiparametric}, Dirichlet process \cite{chib2002semiparametric,karabatsos2012bayesian,roy2018bayesian,oganisian2021practical}, have also been considered for causal inference. We refer interested readers to \cite{linero2022how} for a more detailed review of these methods. 

\subsection{Challenges in High Dimensions} \label{sec::high-dim}
Conducting statistical inference in high dimensions is challenging in general. We differentiate between two high dimensional settings: \emph{(i)} an outcome model with an infinite or a large number of parameters, regardless of the number of covariates, such as nonparametric and semiparametric models, and \emph{(ii)} a large number of covariates. Both settings are increasingly common in causal inference, particularly in studies targeting the CATE.  As discussed earlier, outcome modelling is the primary method in these settings, and Bayesian nonparametric models have become a mainstay of the model choice.

A straightforward application of the standard Bayesian nonparametric priors to outcome modelling is sometimes  inadequate for causal inference, even with low dimensional covariates. An important consideration is that a desirable prior should accurately reflect uncertainty according to the degree of covariate overlap because intuitively the uncertainty of causal estimates should increases as the degree of  overlap decreases. Below, we reproduce a simple example in \cite{papadogeorgou2020discussion} (first constructed by Surya Tokdar) with a single covariate to illustrate this point. 

\begin{figure}[!ht]
  \centering
  \includegraphics[width=0.75\textwidth]{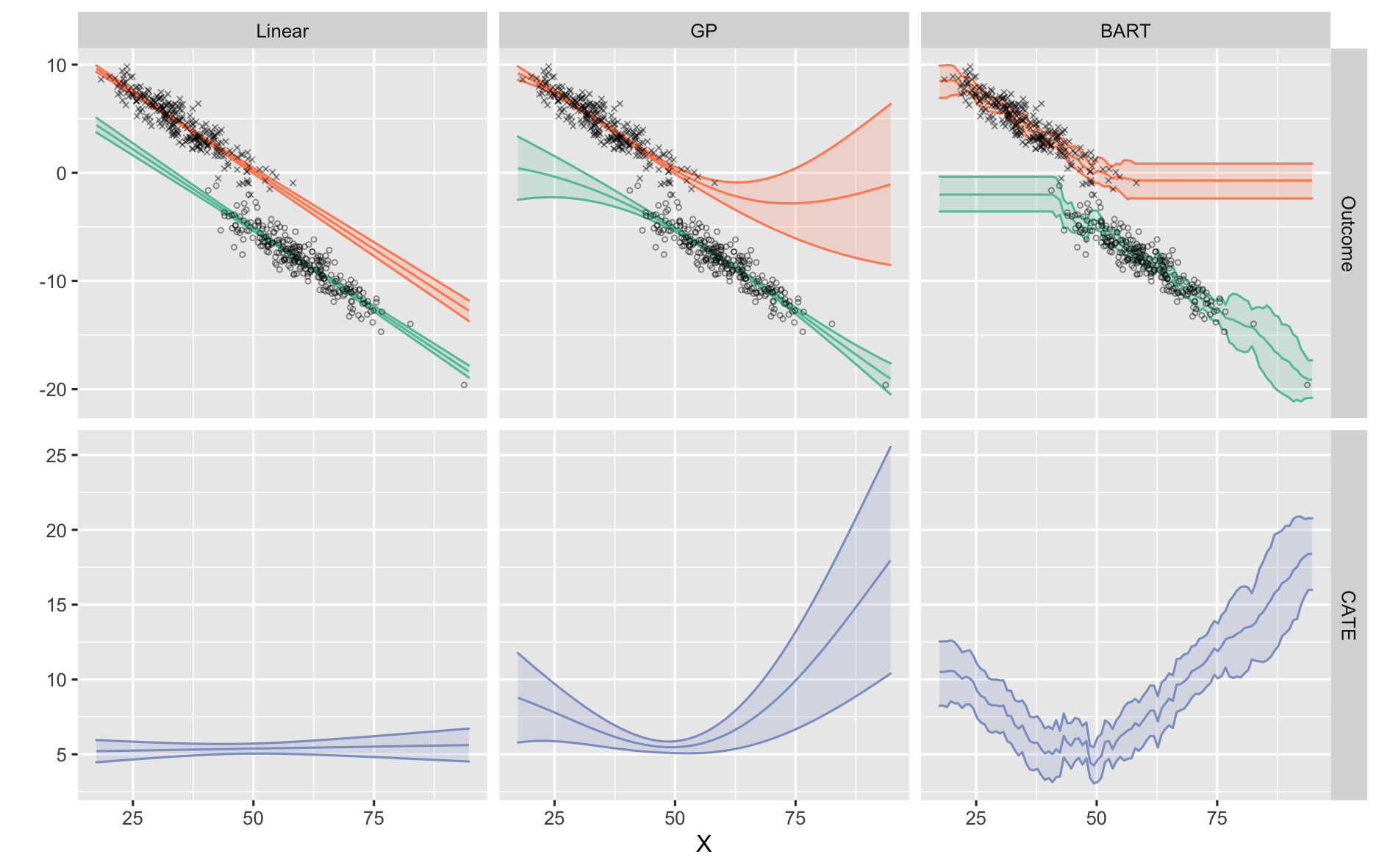}
  \caption{Example 2: Estimates of means of the potential outcomes (upper panel) and the CATE (lower panel) and corresponding uncertainty band as a function of the single covariate by linear model, Gaussian Process, BART, respectively. $\times$: treated units; $\circ$: control units.}
  \label{fig:comparison}
\end{figure}

\textbf{Example 2} [Choice of priors in estimating the CATE] 
Consider a study with $250$ treated and $250$ control units. Each unit has a single covariate $X$ that follows a Gamma distribution with mean $60$ and $35$ in the control ($Z_i=0$) and treatment ($Z_i=1$) group, respectively, and with standard deviation $8$ in both groups. To convey the main message, we consider a true outcome model with constant treatment effects: $Y_i(z)= 10 + 5z-0.3X_i+\epsilon_i$ with $\epsilon_i \sim \mathcal  N(0,1)$, where the CATE $\tau(x)=5$ for all $x$. The scatterplots in the upper panel of Figure \ref{fig:comparison} show that covariate overlap is good between the groups in the middle of the range of $X$ (around 40 to 50), but deteriorates towards the tails of $X$. 

To estimate the CATE, we fit an outcome model separately in each group: $\mu(z,x) = f_z(x) + \epsilon_i$ with $\epsilon_i  \sim \mathcal  N(0, \sigma^2)$. We choose three priors for $f_z(x)$: (1) a linear model $f_z(x) = \alpha_z + \beta_z x$ with a Gaussian prior for the coefficients; (2) a BART prior similar to \cite{hill2011bayesian};  (3) a Gaussian Process prior \cite{rasmussen2003gaussian} with the covariance function specified by a Gaussian kernel with signal-to-noise ratio parameter $\rho$ and inverse-bandwidth parameter $\lambda$: $(f_z(x_1), f_z(x_2), \ldots, f_z(x_N)) ' \sim \mathcal  N(0, \Sigma)$ where $\Sigma_{ij} = \delta^2 \rho^2 \exp\{-\lambda^2\|x_i - x_j\|^2\})$.
Figure \ref{fig:comparison} shows the posterior means of $\mu_z(X)$ and the CATE, with corresponding uncertainty band as a function of $X$. Here we focus on the uncertainty quantification.  In the region of good overlap, all three models lead to similar point and credible interval estimates of the CATE, but marked difference emerges in the region of poor overlap. The linear model appears overconfident in estimating the CATE. Gaussian process trades potential bias with wider credible interval as overlap decreases and produces a more adaptive uncertainty quantification. BART produces shorter error bars than Gaussian Process (but wider than linear models), but the width of the credible interval remains similar regardless of the degree of overlap and thus is over confident in the presence of poor overlap.  $\Box$ 

Example 2 illustrates that, even with low dimensional covariates, standard Bayesian priors can have markedly different operating characteristics when the two groups are poorly overlap, and not all priors can adaptively capture the uncertainty according to the degree of overlap. A primary reason of BART potentially underestimating uncertainty in poor overlap is its lack of smoothness, in contrast of Gaussian process. Nonetheless, such a problem can be mitigated by soft decision trees as in \cite{linero2018bayesian}.

When there are a large number of covariates, the Bayesian paradigm usually achieves regularization through sparsity-inducing priors for the outcome model, such as the spike-and-slab prior \cite{antonelli2019highdim}, the Bayesian LASSO \cite{park2008bayesian}, as well as the model averaging techniques \cite{raftery1997bayesian, Wang12, zigler2016central}. The use of these methods in causal inference is surveyed in \cite{oganisian2021practical, linero2022how}. High dimensional covariates pose additional complications to Bayesian causal inference. Ritov et al. \cite{ritov2014bayesian} pointed out that nonparametric estimates often have slow convergence rates in this regime, which translates into poor finite-sample performance. A central challenge in causal inference is that covariate overlap is rapidly diminishing as the covariates dimension increases, violating the overlap assumption that underpins standard causal analysis \cite{damour2021overlap}. Lack of overlap exacerbates the usual inferential challenges---such as sparsity and slow convergence---in high dimensional analysis. Even if we assume linear outcome models, we must carefully specify the priors on the regression coefficients. For example, Hahn et al. \cite{hahn2018regularization} showed that standard Bayesian regularization on the nuisance parameters may indirectly regularize important causal parameters and thus induce bias, namely the \emph{regularization induced confounding}. This issue was rigorously investigated in \cite{linero2021nonparametric}. Specifically,  Linero  \cite{linero2021nonparametric} defines the selection bias as $\delta_z=\bE(Y_i\mid Z_i=z)-\bE\{Y_i(z)\}$, and showed that, under the seemingly innocuous prior independence assumption \ref{ass::prior-independence}, many Bayesian regularization priors would \emph{a priori} induce the selection bias $\delta_z$ to sharply concentrate around zero as the number of covariates, $p$, increases, to the extent that no amount of data would overcome such a bias. This implies that  Assumption \ref{ass::prior-independence} effectively acts as a strongly informative prior as $p$ increases. Such a phenomenon is referred to as \emph{prior dogmaticism} and is the Bayesian analogue of the aforementioned problem in Ritov et al. This line of research highlighted the importance of incorporating the propensity score in Bayesian causal inference \cite{harmeling2007bayesian,sims2012robins,hahn2018regularization, linero2021nonparametric}, which echos the insights from the Frequentist \emph{double machine learning} method \cite{chernozhukov2017double, antonelli2022causal}. Specifically, regularized propensity score model or outcome model alone would not be sufficient for valid causal inference, but combining two would achieve desirable convergence rate and finite sample performance in high-dimensional causal analysis. 

\section{The Role of the Propensity Score}
\label{sec::PSrole}

A major debate in Bayesian causal inference is the role of the propensity score, which characterizes the assignment mechanism.  On the one hand, as shown in Section \ref{sec::BayesianCI_general}, under Assumptions \ref{ass::ignorability} and \ref{ass::prior-independence}, the propensity score drops out from the likelihood and thus its value appears to be irrelevant in Bayesian causal inference, which seemingly only involves the outcome model and thus the analysis stage. On the other hand, Section \ref{sec::primitives} shows that the propensity score is ubiquitous in the Frequestist approach to causal inference, e.g. in constructing weighting, matching, and doubly-robust estimators. Regardless of the mode of inference, the propensity score is essential in ensuring overlap and balance in the design stage of an observational study, which consequently reduces the sensitivity to the outcome model specification. Such sensitivity reduction is key to robust Bayesian causal inference, which is primarily based on outcome modelling. The literature has recognized the importance of incorporating the propensity scores into Bayesian causal inference, either in the design or the analysis stage, but there is no consensus on how to proceed. Below we review three existing strategies.
 
\subsection{Include the Propensity Score as a Covariate in the Outcome Model} \label{sec::PSrole-pscovariate}
The propensity score was first proposed to be included as the \emph{only} covariate in a Bayesian outcome model under ignorability, which would reduce the model complexity \cite{Rubin85}. However, as later pointed out by Zigler \cite{zigler2016central}: $\Pr\{ Y(z)\mid X,e(X)\} = \Pr\{ Y(z) \mid X \}  \neq\Pr\{ Y(z) \mid e(X) \} $. So using the propensity score as the single covariate in the outcome model would not lead to the target outcome distribution $\Pr\{ Y(z)\mid X \} $, but using it as an additional covariate, i.e. specifying a model $\mu(z, x)= \mu(z, x, e(x))$, would. This specification is effectively conducting an outcome regression at each value of the propensity score, and thus can be viewed as a smoothed version of combining propensity score stratification and outcome modelling. In a sense, this specification provides a Bayesian analogue of the double robustness \cite{Zigler13,papadogeorgou2020discussion}. On the one hand, when the outcome model is correctly specified, $\mu(Z_i,X_i,e(X_i))$ reduces to $\mu(Z_i,X_i)$ because $e(X_i)$ is a function of $X_i$ and thus is redundant regardless of its specification. On the other hand, when the outcome model is misspecified but the propensity model is correctly specified, the results are robust to the outcome model specification because the treatment and control groups are approximately balanced in covariates within each stratum of the propensity score. Various reparametrization has been proposed. One example is to specify $\mu(z, x, e(x)) = g_1(x, e(x)) + g_2(z, x)$, with $g_1(\cdot)$ being a nonparametric model and $g_2(\cdot)$ being a parametric model. Little \cite{little2004robust} adopted a penalized spline model of $e(x)$ for $g_1(\cdot)$.  In the aforementioned \emph{Bayesian Causal Forest}, Hahn et al. \cite{hahn2020bayesian} imposed a separate BART model for $g_1(\cdot)$ and $g_2(\cdot)$, and demonstrated that adding the propensity score as an additional predictor in $g_1$ significantly improves the empirical estimation of the CATE.    

This strategy is usually implemented in two stages: first estimate the propensity score as $\widehat{e}(X)$ and then plug it into the Bayesian outcome model $\mu(Z, X, \widehat{e}(X))$. Such a two-stage procedure is not dogmatically Bayesian, which generically refers to the procedure of specifying a model with parameters and prior distributions of these parameters and then use the Bayes theorem to obtain the posterior distributions of the parameters. A direct consequence is that this procedure may not properly propagate the uncertainty of estimating the propensity score in the outcome model \cite{Zigler13}. A dogmatic Bayesian approach would jointly model $e(X; \theta_Z)$ and $\mu(Z, X, e(X); \theta_Y)$ and draw posterior inference of $\theta_Z$ and $\theta_Y$ simultaneously \cite{McCandless2009PS}. However, when the outcome model is misspecified, the joint-modelling approach would introduce a \emph{feedback} problem, that is, the fit of the outcome model would inform the estimation of the propensity scores. This violates the unconfoundedness assumption, distorts the balancing property of the propensity score, and consequently leads to biased estimate of causal effects. A suggested remedy is to first fit a Bayesian model for $e$ and then plug the posterior predictive draws of $e$ into the outcome \cite{ZiglerDominici14}. Such a two-stage procedure is still not dogmatically Bayesian, but provides more robust posterior inference to model misspecification empirically.

However, adding propensity score into the outcome model is controversial conceptually, because the outcome model reflects the nature of the generating process of the potential outcomes, which arguably should not depend on how the treatment is assigned \cite{robins2015bayesian}.

\subsection{Dependent Priors}
The Bayesian causal inference outlined in Section \ref{sec::BayesianCI_general} rests on the prior independence assumption \ref{ass::prior-independence}, without which the propensity score model can not be ignored from the likelihood. But this assumption is not always plausible in real applications. Various priors that do not rely on this assumption have been proposed \cite{harmeling2007bayesian,sims2012robins,ritov2014bayesian,ray2020semiparametric}.  Below we show two simple examples. 

The first example is due to \cite{Wang12} and is designed for simultaneous variable selection for the propensity score and outcome models. Specifically, assume a logistic propensity score model logit$\{ \Pr(Z_i = 1\mid X_i) \} = \alpha'X_i$ and a linear outcome model $Y_i\mid Z_i, X_i \sim \mathcal N (\tau Z_i + \beta' X_i, \sigma^2)$. 
Assume each of the $j$th  component of the coefficients, $\alpha_j$, follow the spike-and-slab prior \cite{George97}:   
$\alpha_j\mid \gamma^{\alpha}_j \sim (1-\gamma^{\alpha}_j)I_0+ \gamma^{\alpha}_j \mathcal N(0, \sigma_{\alpha}^2)$, where $\gamma^{\alpha}_j$ is a latent indicator of whether $X_j$ is included in the model and $I_0$ denotes the point mass at $0$. A similar spike-and-slab prior is assumed for the coefficients of the outcome model with a latent inclusion indicator $\gamma^{\beta}_j$: $\beta_j \mid \gamma^{\beta}_j \sim (1-\gamma^{\beta}_j)I_0+ \gamma^{\beta}_j \mathcal N(0, \sigma_{\beta}^2)$. Then assume the probability of the events $\{\gamma^{\alpha}_j=0\}$ and $\{\gamma^{\beta}_j =0\}$ are dependent \emph{a priori}:
$ {\Pr(\gamma^{\beta}_j=1\mid \gamma^{\alpha}_j=1)}/{\Pr(\gamma^{\beta}_j=0\mid \gamma^{\alpha}_j=1)} =\omega$, 
where $\omega \in [1, \infty)$ is a dependence hyperparameter that controls the prior odds of including $X_j$ into the outcome model when it is included in the propensity score model. Larger $\omega$ implies stronger prior dependence between the variable selection in the two models.  

The second example is due to \cite{little2004model}. Assume $Y_i(1)\mid X_i \sim \mathcal{N}(\mu_1, \sigma_1^2 e(X_i))$ and $Y_i(0)\mid X_i \sim \mathcal{N}(\mu_0, \sigma_0^2 (1-e(X_i)))$, with flat priors on $\mu_1$ and $\mu_0$. If the propensity scores are known, the posterior mean of the PATE equals the Haj\'ek version of the IPW estimator: $\widehat{\tau}^{\text{hajek}}  =  \sum_{i=1}^N \{Z_iY_i/e(X_i)\} /  \sum_{i=1}^N \{Z_i/e(X_i)\}  - \sum_{i=1}^N \{(1-Z_i)Y_i/(1-e(X_i))\} / \sum_{i=1}^N \{(1-Z_i)/(1-e(X_i))\} $. If the propensity scores are unknown, then the posterior mean of the PATE is closely related to $\widehat{\tau}^{\text{hajek}}$ averaged over the posterior predictive distribution of the propensity scores. This strategy simply includes the propensity scores in the outcome model, but somewhat unusually in the conditional variances, rather than the conditional means of the potential outcomes.

Carefully designed dependent priors often achieve desirable finite sample results and are more reasonable in real world studies. However, specification of such priors is case-dependent, and there is no general solution.

\subsection{Posterior Predictive Inference}

A general, albeit not dogmatically Bayesian, strategy is to specify both a propensity score model $e(X_i; \theta_Z)$ and an outcome model $\{ \mu_1(X_i; \theta_Y), \mu_0(X_i; \theta_Y) \}$, and obtain posterior draws of $e(X_i; \theta_Z)$ and $\{ \mu_1(X_i; \theta_Y), \mu_0(X_i; \theta_Y) \}$ from their respective posterior predictive distributions, and then plug these posterior draws into the doubly-robust estimator $\hattaudr$ \cite{saarela2016bayesian, antonelli2022causal}. A variance estimator of the resulting estimator $\hattaudr$ is given in \cite{antonelli2022causal}.
In the same vein, Ding and Guo \cite{ding2022posterior} incorporated the propensity score in Bayesian posterior predictive $p$-value. For the model with the Fisher's sharp null hypothesis of no causal effect for any units whatsoever (i.e., $Y_i(1)=Y_i(0)$ for all $i$), the procedure in \cite{ding2022posterior} is equivalent to the Fisher randomization test averaged over the posterior predictive distribution of the propensity score.  Simulations in \cite{ding2022posterior} show the advantages of the Bayesian $p$-value compared to the Frequentist analogue.  This perspective offers a straightforward and flexible strategy to integrate Bayesian modelling and common Frequentist procedures for causal inference and enables proper uncertainty quantification.  



Besides the above three strategies, another general approach is through the aforementioned Bayesian bootstrap, which can be used to simulate the posterior distribution of any parameter that can be formulated as M-estimation or estimating equation \cite{Chamberlain2003, Lyddon2019}. As special cases, because the IPW estimator $\widehat\tau^\text{ipw}$ and the doubly-robust estimator $\widehat\tau^\text{dr}$--both involving the propensity scores---are both solutions to estimating equations, they can be naturally combined with the Bayesian bootstrap to devise a Bayesian version. However, such an approach may be guilty of ``Bayesian for the sake of being Bayesian", and their methodological and practical value compared to competing methods is unclear. 


\section{Sensitivity Analysis in Observational Studies} \label{sec::SA}

Unconfoundedness is a central assumption for causal inference. It holds by design in randomized experiments. However, its validity is fundamentally untestable in observational studies. Therefore, it is of great importance to assess the sensitivity of the results with respect to unmeasured confounding in any observational study. Such procedures are broadly called {\it sensitivity analysis}. Different classes of sensitivity analysis methods are characterized by the specific parametrization of confounding. Below we review the two most popular classes. 

\subsection{Parametrization involving distributions with unmeasured confounders} \label{sec::SA-confounder}

The first parametrization used for sensitivity analysis is motivated by the intuition that a hidden confounder may completely explain away the  association between the treatment and the outcome even after adjusting for observed covariates. In a historic debate, Fisher \cite{fisher1957dangers} hypothesized that the strong association between cigarette smoking and lung cancer might be due to a hidden genetic factor as their ``common cause" or confounder. Cornfield et al. \cite{cornfield1959smoking} derived an inequality showing that to explain away the observed association, the association between the unmeasured confounder and cigarette smoking must be larger than or equal to the association between cigarette smoking and lung cancer. Their work helped to initiate the field of sensitivity analysis. 

Let $U$ denote an unmeasured confounder and assume that unconfoundedness holds conditional on  $(X, U)$: $Z\ci \{Y(0),Y(1)\} \mid X, U$. The joint distribution of all variables factorizes into 
\begin{equation}\label{eq::factorization-sa}
     \Pr\{ Y(1), Y(0), Z, X, U \}  = \Pr\{ Y(1), Y(0)\mid X, U\}  \cdot \Pr(Z \mid X, U) \cdot \Pr(U\mid X) \cdot \Pr(X).
\end{equation}
Under the factorization \eqref{eq::factorization-sa}, sensitivity analysis requires to specify the models for $\Pr\{ Y(1), Y(0)\mid X, U\} $, $\Pr(Z \mid X, U)$, and $\Pr(U\mid X)$. In the special case of a binary $Z$, a binary $Y$ and a discrete $X$ (which can be thought as a stratified propensity score), Rosenbaum and Rubin \cite{rosenbaum1983assessing} assumed a logistic model for $Y$ given $(Z,X,U)$, a logistic model for $Z$ given $(X, U)$, and a Bernoulli distribution for $U$, and treated the logistic regression coefficients of $U$  and the probability parameter of $U$ as the sensitivity parameters. They integrated out $U$ in the complete-data likelihood and obtained the maximum likelihood estimates of $\pate$ over a plausible range of values of the sensitivity parameters. This method has been extended to more general settings in the Frequentist fashion in \cite{imbens2003sensitivity,ichino2008temporary}. The Bayesian analogue of \cite{rosenbaum1983assessing} is straightforward and can leverage the data  augmentation algorithm to impute $U$ to simplify the computation. Dorie et al. \cite{dorie2016SA} extended this method to impose a Bayesian semiparametric model with a BART component for $\Pr\{ Y(1), Y(0)\mid X, U\} $ to allow for model flexibility.

As an extension of \cite{cornfield1959smoking}, Ding and VanderWeele \cite{ding2016sensitivity} treated the treatment-confounder ($Z$ and $U$) and outcome-confounder ($Y$ and $U$) associations as two sensitivity parameters, and derived analytical thresholds for them in order to explain away the observed treatment-outcome ($Z$ and $Y$) association. Based on that theory, VanderWeele and Ding \cite{vanderweele2017sensitivity} further simplified by assuming the two associations to be the same and called the resulting threshold the \emph{E-value}, as a measure of robustness of the causal conclusions with respect to unmeasured confounding. The E-value framework is model-free because it avoids modelling assumptions with $U$; it also avoids repeating the analysis over a range of sensitivity parameters as in the competing methods and thus is simple to calculate.

\subsection{Parametrization involving distributions of potential outcomes}\label{sec::SA-outcome}
        
The second parametrization is motivated by an alternative mathematical expression of the unconfoundedness assumption: $\Pr\{ Y(z) \mid Z=1, X \} = \Pr\{ Y(z) \mid Z=0, X \}$ for $z=0,1$, representing the fact that the units in the two randomized arms are comparable in terms of potential outcomes. This class of sensitivity analysis is based on sensitivity parameters that directly represent the difference between the distributions $\Pr\{ Y(z) \mid Z=1, X \}$ and $\Pr\{ Y(z) \mid Z=0, X \}$ instead of modelling the difference with an unobserved $U$. This is implemented in the context of time-varying treatments (see Section \ref{sec::extensions}\ref{sec::time-varying}) and Frequentist semiparametric estimation \cite{robins1999association}. Franks et al. \cite{franks2019flexible} pointed out the importance of distinguishing between model fit and sensitivity to unconfoundedness: the former involves identifiable parameters (e.g. the parameters in the model of the marginal distributions of the outcome $\Pr\{Y_i(z)\mid Z_i=z, X\}$) whereas the latter involves unidentifiable parameters (e.g. the association between $Y_i(1)$ and $Y_i(0)$). The merit of this parametrization is apparent in this perspective because it separates identifiable and unidentifiable parameters. Franks et al. \cite{franks2019flexible} proceeded under the Bayesian paradigm and used a copula---parameters of which are the sensitivity parameters---to connect the two identifiable marginal distributions.

A related branch of sensitivity analysis is Rosenbaum's bounds  \cite{rosenbaum2002observational}. His original formulation takes the association between $Z$ and the potential outcomes conditional on observed $X$, denoted by $\Gamma$, as the sole sensitivity parameter for quantifying unmeasured confounding.
He has also made connections to the parametrization in Section \ref{sec::SA}\ref{sec::SA-confounder} \cite{rosenbaum2009Amplification}. Starting with a matched sample to mimic a randomized  matched-pairs experiment, one can then repeat the Fisher randomization test on the sharp null hypothesis of no treatment effect given a range of $\Gamma$ values, and find the threshold of $\Gamma$ at which the $p$-value of the test changes from significant to insignificant. This approach was later generalized to derive the bounds of a given estimator under different $\Gamma$ values. Grounded in Fisherian randomization inference, Rosenbaum's framework does not have a natural Bayesian analogue.

Besides the above two classes, there are numerous other approaches of sensitivity analysis based on alternative parameterization of unmeasured confounding. However, a common criticism of various sensitivity analysis is that, in order to assess the consequence of the untestable unconfoundedness assumption, one has to make even more untestable assumptions, e.g. specifying models involving $U$. Moreover, sensitivity analysis, after all, is a secondary analysis in causal studies, and thus simple implementation and intuitive interpretation is much desired. The considerations underpin the dominance of the E-value method over other methods in practice, particularly in medicine and public health.

\section{Complex Assignment Mechanisms}\label{sec::extensions}
So far we have discussed the simplest causal setting of an ignorable treatment at one time point. The basic formulation can be extended to many more complex assignment mechanisms. There are also many popular quasi-experimental designs that rely on identification strategies alternative to ignorability, e.g. regression discontinuity designs, difference-in-differences, synthetic controls. These designs are widely used in socioeconomic applications. Due to the space limit, below we will only briefly review two important extensions and refer interested readers to \cite{angrist2009mostly} for a review of quasi-experimental designs and related econometric methods.

\subsection{From Instrumental Variable to Principal Stratification}
\label{sec::principalstratification}
Instrumental variable (IV) is one of the most important techniques for causal inference in economics and social sciences. IVs are used in settings where dependence of the assignment on the potential outcomes cannot plausibly be ruled out, even conditional on observed covariates. An IV is a variable that provides a source of \emph{exogenous} (or unconfounded) variation that helps identifying causal effects. IV methods are based on a set of assumptions alternative to ignorability. Specifically, an IV satisfies three conditions: \emph{(i)} it occurs before a treatment; \emph{(ii)} it is independent of the treatment-outcome confounding; and \emph{(iii)} it  affects the outcome only through its (non-zero) effects on the treatment assignment. Finding a valid IV is challenging in observational studies and many clever natural experiments have been identified \cite{angrist2009mostly}. Given a valid IV, one can extract the causal effects of the treatment on an outcome by a two-stage least squares (2SLS) estimator: first, fit a linear regression of the treatment on the IV; second, fit a linear regression of the outcome on the fitted value of the treatment from the first stage, the coefficient of which is taken as the causal effect of the treatment on the outcome. Covariates can be added in both stages.  

The IV method has been developed within the structural equation model framework (see \cite{angrist2009mostly} for a review), and the 2SLS IV estimator may not correspond to a causal effect within the potential outcomes framework except for a few special cases. In a landmark paper, Angrist et al. \cite{Angrist96} connects IV to the potential outcomes framework in the setting of randomized experiments with binary treatment and all-or-nothing compliance, with the initial random assignment playing the role of an IV. But many questions remain on the connection between the IV method and the potential outcomes framework in more general settings. 
Below we will describe the special setting of Angrist et al. \cite{Angrist96}.

We introduce some new notation. For unit $i$, let $Z_i$ be the randomly assigned treatment ($1$ for the treatment and $0$ for the control), and $W_i$ be the actual treatment status ($1$ for the treatment and $0$ for the control). When $Z_i \neq W_i$, noncompliance arises. Because $W_i$ occurs post assignment, it has two potential values, $W_i(0)$ and $W_i(1)$, with $W_i=W_i(Z_i)$. As before, the outcome $Y_i$ has two potential outcomes, $Y_i(0)$ and $Y_i(1)$. 
Based on their joint potential status of the actual treatment $U_i = (W_i(1), W_i(0))$, the units fall into four compliance types: compliers $U_i=(1,0) = \co$, never-takers $U_i=(0,0) = \nt$,   always-takers $U_i=(1,1) =\at$, and defiers $U_i=(0,1) = \df$ \cite{Angrist96}.  A key property of $U_i$ is that it is not affected by the treatment assignment, and thus can be regarded as a pre-treatment characteristic. Therefore, comparisons of $Y_i(1)$ and $Y_i(0)$ within the stratum of $U_i$ have standard subgroup causal interpretations: $\tau_u = \bE\{ Y_i(1) - Y_i(0) \mid U_i  =u \}$, 
for $u=\nt,\co,\at,\df$; $\tau_u$ are later called {\it principal causal effects}. The conventional causal estimand in clinical trials is the intention-to-treat effect that ignores the compliance information, which is the weighted average of the four stratum-specific effects:
$ \bE\{ Y_i(1) - Y_i(0) \} = \sum_{u}  \pi_u \tau_u,$
where $\pi_u  = \Pr( U_i = u) $ is the proportion of the stratum $u$. The intention-to-treat effect measures the effect of the assignment instead of the actual treatment.

Due to the fundamental problem of causal inference, individual compliance stratum $U_i$ is not observed. So the principal causal effects are non-identifiable without  additional assumptions. Besides randomization of $Z_i$, Angrist et al. \cite{Angrist96} make two additional assumptions: \emph{(i)} monotonicity: $W_i(1) \geq W_i(0)$, and \emph{(ii)} exclusion restriction: $Y_i(1) = Y_i(0)$ whenever $W_i(1) = W_i(0)$. Monotonicity rules out defiers, and exclusion restriction imposes that the assignment has zero effects among never-takers and always-takers. Then the complier average causal effect  is identified by 
$$
\tau_{\co}
\equiv \bE\{ Y(1) - Y(0) \mid U = \co   \}
= \frac{  \bE(Y\mid Z=1)-\bE(Y\mid Z=0) }{  \bE(W\mid Z=1)-\bE(W\mid Z=0) } ,
$$ 
which is exactly the probability limit of the two-stage least squares estimator \cite{Angrist96}. Because under monotonicity, only compliers' actual treatments are affected by the assignment, $\tau_{\co}$ can be interpreted as the effect of the treatment.

We now describe the Bayesian inference of the IV setup, first outlined in \cite{ImbensRubin97}. Without additional assumptions, the observed cells of $Z$ and $W$ consist of a mixture of units from more than one stratum. For example, the units who are assigned to the treatment arm and took the treatment ($Z=1, W=1$) can be either always-takers or compliers. One must disentangle the causal effects for different compliance types from observed data. Therefore, model-based inference here resembles that of a mixture model. In Bayesian analysis, it is natural to impute the missing label $U_i$ under some model assumptions. Specifically, six quantities are now associated with each unit, $\{Y_i(1)$, $Y_i(0)$, $W_i(1)$, $W_i(0)$, $X_i$, $Z_i\}$, four of which, $\{ Y_i^\obs =  Y_i =Y_i(Z_i), W_i^\obs =  W_i =W_i(Z_i), Z_i, X_i\}$, are observed and the remaining two, $\{ Y_i^\mis =  Y_i(1-Z_i), W_i^\mis =  W_i(1-Z_i)\}$, are unobserved. Assume the joint distribution of these random variables of all units is governed by a parameter $\theta$, conditional on which the random variables for each unit are \emph{iid}. We assume unconfoundedness $\Pr\{Z_i=1 \mid X_i,  W_i(1), W_i(0), Y_i(1), Y_i(0)\}=\Pr(Z_i=1\mid X_i)$, and impose a prior distribution $p(\theta)$. Then the joint posterior distribution of $\theta$ and the missing potential outcomes is proportional to the complete-data likelihood as follows:
\begin{equation}
    \label{eq::posterior-IV}
p(\theta) \prod_{i=1}^N \Pr\{Y_i(0),Y_i(1)\mid U_i, X_i;  \theta_Y\}\Pr(U_i \mid X_i;  \theta_U)  \Pr (X_i \mid  \theta_X ). 
\end{equation}
Without covariates, posterior inference of $\tau_u$ is straightforward because it is a function of $\theta_Y$ (see Example 3 below). With covariates, we can condition on them and focus on a MATE estimand $\tau_u^\textsc{m} = N^{-1} \sum_{i=1}^N \bE\{ Y_i(1) - Y_i(0) \mid U_i=u, X_i \} $.
The formula \eqref{eq::posterior-IV} suggests that we need to specify two models for inferring $\tau_u^\textsc{m}$: \emph{(i)} the compliance type model, $\Pr \left(U_i\mid X_i;  \theta_U\right)$, and \emph{(ii)} the outcome model, $\Pr\{Y_i(0),Y_i(1)\mid U_i, X_i; \theta_Y\}$. For example, we can specify a multinomial logistic regression for $U_i$ and a generalized linear model for $Y_i$ \cite{ImbensRubin97, hirano2000}.

Using the same arguments as in Section \ref{sec::BayesianCI_general}, to infer population and mixed estimands, we only need to specify two marginal outcome models for $Y_i(1)$ and $Y_i(0)$ instead of a joint model, and do not need to impute the missing potential outcomes $Y^{\mis}$. But we do need to impute the latent $U_i$, or, equivalently, the missing intermediate variable $W^{\mis}$. We can simulate the joint posterior distribution $\Pr(\theta, \bW^{\mis}\mid\bY^{\obs}, \bW^{\obs},\bZ, \bX)$ by iteratively imputing the missing $\bU$ from $\Pr(\bW^{\mis} \mid\bY^{\obs}, \bW^{\obs}, \bZ, \bX, \theta)$ and updating the posterior distribution of $\theta$ from $\Pr(\theta\mid\bY^{\obs}, \bW^{\obs}, \bW^{\mis}, \bZ, \bX)$.

Below we illustrate the Bayesian procedure via a simple example of the IV approach.

{\bfseries Example 3} [Randomized experiment with one-sided noncompliance]
Consider a randomized experiment with a binary outcome, where control units have no access to the treatment, i.e. $W_i(0)=0$ for all units. Therefore, we only have two strata: $U_i = \co$ with $W_i(1)=1$ and $U_i = \nt$ with $W_i(1)=0$, respectively, with $\pi_\co+\pi_\nt=1$. Assume $Y_i(z)\mid U_i=\co \sim \textup{Bern}(p_{\co,z})$ for $z=0,1$, and   $Y_i(1)=Y_i(0)\mid U_i=\nt \sim \textup{Bern}(p_{\nt})$. So $\tau_{\co} = p_{\co,1}- p_{\co,0}$. For simplicity, we impose conjugate priors on the parameters: $\pi_\co , p_{\co,1}, p_{\co,0}, p_{\nt}$ are \emph{iid} Beta$(1/2, 1/2)$. To sample the posterior distributions, the key is to impute the missing $U_i$'s given the observed data. If $Z_i=1$, then $W_i=1$ implies $U_i=\co$ and $W_i=0$ implies $U_i=\nt$, respectively. If $Z_i=0$, then $W_i=0$ and $U_i$ is latent. For units with $(Z_i=0,W_i=0)$, we can impute $U_i = \co$ with probability 
$$
\frac{  \pi_\co \cdot p_{\co,0}^{Y_i} (1-p_{\co,0})^{1-Y_i}  }
{  \pi_\co \cdot p_{\co,0}^{Y_i} (1-p_{\co,0})^{1-Y_i} 
+  \pi_\nt \cdot p_{\nt}^{Y_i} (1-p_{\nt})^{1-Y_i}  }
$$
and $U_i = \nt$ with the rest probability. 
With the imputed $U_i$'s, we can sample the parameters from standard Beta posteriors: \emph{(i)} sample $\pi_\co$ from Beta$(1/2+\sum_{i=1}^N \one(U_i=\co) , 1/2+\sum_{i=1}^N \one(U_i=\nt))$ and obtain $\pi_\nt = 1-\pi_\co$, \emph{(ii)} sample $p_{\co,1}$ from Beta$(1/2+\sum_{i=1}^N Z_i\one(U_i=\co)Y_i, 1/2+\sum_{i=1}^N Z_i\one(U_i=\co)(1-Y_i))$, \emph{(iii)} sample $p_{\co,0}$ from Beta$(1/2+\sum_{i=1}^N (1-Z_i)\one(U_i=\co)Y_i, 1/2+\sum_{i=1}^N (1-Z_i)\one(U_i=\co)(1-Y_i))$, and (iv) sample $p_\nt$ from Beta$(1/2+\sum_{i=1}^N \one(U_i=\nt) Y_i , 1/2+\sum_{i=1}^N \one(U_i=\nt) (1-Y_i))$. We iterate until convergence and obtain the posterior distribution of $\tau_{\co} = p_{\co,1}- p_{\co,0}$. Imbens and Rubin  \cite{ImbensRubin97} provided more detailed discussions. $\Box$

Frangakis and Rubin \cite{FrangakisRubin02} generalized the IV approach to principal stratification, a unified framework for causal inference with post-treatment confounding. In the simplest scenario, a post-treatment confounded variable lies in the causal pathway between the treatment and the outcome; it cannot be adjusted in the same fashion as a pre-treatment covariate in causal inference.  A principal stratification with respect to a post-treatment variable is the classification of units based on the joint potential values of the post-treatment variable, and the stratum-specific effects are called \emph{principal causal effects}, of which $\tau_{\co}$ is a special case.  The post-treatment variable setting includes a wide range of examples. For instance, in the noncompliance setting, the ``treatment" is the randomized treatment assignment, the ``post-treatment" variable is the actual treatment received, and the compliance types are the principal strata \cite{ImbensRubin97, hirano2000, FrangakisRubinZhou2002, mealli2013using}. Zeng et al. \cite{zengliding2020jrssa} connects  principal stratification to the local IV method with a continuous IV and binary treatment  \cite{heckman1999local}. Other examples include censoring due to death \cite{Zhang08}, surrogate endpoints \cite{Gilbert08}, regression discontinuity designs \cite{li2015evaluating}, time-varying treatments \cite{ricciardi2020bayesian}, and many more. The choices of target strata and thus estimands, interpretations, and identifying assumptions depend on specific applications, details of which are omitted here.

\subsection{Time-Varying Treatment and Confounding}
\label{sec::time-varying}

In real world situations, subjects often receive treatments sequentially at multiple time points, and the treatment assignment at each time is affected by both baseline and time-varying confounders as well as previous treatments \cite{Robins:1986, RHB:2000}. Such settings are referred to as time-varying, or sequential, or longitudinal treatments. 

Consider a study where treatments are assigned at $T$ time points. Let $Z_{it}$ denote the treatment at time $t$ for unit $i$ $(i=1,\ldots, N; t=1,\ldots, T)$. At baseline ($t=0$), each unit $i$ has time-invariant covariates $L_{i0}$ measured; then after the treatment assignment at time $t-1$ and prior to the assignment at time $t$, a set of time-varying confounders $L_{i, t-1}$ are measured, which include the intermediate measurements of the final outcome and the covariates that are affected by the previous treatments. For example, in a cancer study, baseline covariates can include sex, age, race, and time-varying confounders can include intermediate cancer progression and other clinical traits such as blood pressure measured prior to the next treatment. Denote the observed and hypothetical treatment sequence of length $t$ by $\bar{Z}_{it} = (Z_{i1},..., Z_{it})$ and $\bar{z}_t=(z_1,...,z_t)$, respectively, and the sequence of time-varying confounders by $\bar{L}_{it} = (L_{i0}, L_{i1},..., L_{it})$. For each  $\bar{z}_T$, there is a potential outcome $Y(\barz_T)$. The final observed outcome $Y_{i}=Y_i(\bar{Z}_{iT})$, corresponding to the entire observed treatment sequence, is measured after treatment assigned at $T$. A common causal estimand is the marginal effect comparing two pre-specified treatment sequences, $\barz, \barz' \in \{ 0,1 \}^T$:  $\tau_{\barz_T,\barz_T'}=\bE\{ Y_i(\barz_T)-Y_i(\barz_T') \} $. For simplicity, below we drop the subscript $i$.

The central question to causal inference with sequential treatments is the role of the time-varying confounders $L_t$ in the assignment mechanism. These variables are affected by the previous treatments and also affect the future treatment assignment and outcome. Much of the literature assumes a \emph{sequentially ignorable} assignment mechanism \cite{Robins:1986}, that is, the treatment at each time is unconfounded conditional on the observed history, which consists of past treatments $\bar{Z}_{t-1}$ and time-varying confounders $\bar{L}_{t-1}$, as stated below.

\begin{ass}\label{ass:SI_AM} \textbf(Sequential Ignorability). 
$\Pr\{ Z_{t} \mid \bar{Z}_{t-1}, \bar{L}_{t-1}, Y(\barz_t) \text{ for all } \barz_t \}  = \Pr(Z_{t} \mid \bar{Z}_{t-1}, \bar{L}_{t-1})$ for $t=1,\ldots, T.$
\end{ass}

A full Bayesian approach to time-varying treatments \cite{Zajonc12} would specify a joint model for treatment assignment $Z_t$ and time-varying confounders $L_t$ at all time points as well as all the potential outcomes $Y(\barz_T) $, and then derive the posterior predictive distributions of the missing potential outcomes and thus of the estimands. This procedure is a straightforward extension from the structure introduced in Section \ref{sec::BayesianCI_general}.  
However, the joint modelling approach is rarely used because it quickly becomes intractable as the time $T$ and the number of time-varying confounders increases. 

Instead, most of the Bayesian methods are grounded in the g-computation. Under sequential ignorability, the average potential outcome $\bE\{ Y_i(\barz_T) \} $ is identified from the observed data via the g-formula \cite{Robins:1986}:
\begin{eqnarray}
\bE\{ Y_i(\barz_T) \}  &=& \sum_{L_0, L_1,...,L_{T-1}}  \bE(Y \mid \barZ_T=\barz_T, \bar{L}_{T-1}) \nonumber  \\
&&\quad \cdot  \Pr(L_{T-1} \mid \barZ_{T-1}=\barz_{T-1,} \bar{L}_{T-2})  \cdots \Pr(L_1 \mid Z_1 =z_1, L_0) \cdot \Pr(L_0). \label{eq::g-formula} 
\end{eqnarray}
To operationalize the g-formula, we can specify models for all the components of  \eqref{eq::g-formula}, including an outcome model $ \Pr(Y \mid  \barZ_T=\barz_T, \bar{L}_{T-1})$ and a model for the time-varying confounders $L_t$ at each time $t$, $\Pr(L_t \mid \barZ_t=\barz_t, \bar{L}_{t-1})$. The g-formula is in essence an extension of the outcome regression approach to time-varying treatments.  The Bayesian version of the g-computation would specify a Bayesian model for each component in the g-formula \eqref{eq::g-formula} and then combine the posterior draws of the parameters to obtain the posterior distribution of the estimands. 
Below we present an illustrative example of Bayesian g-computation due to  \cite{Gustafson15}. 

{\bfseries Example 4} [Bayesian g-computation with two periods]
Consider the simplest possible scenario with two time periods, binary covariates and a binary outcome. Let $L_0$ be a binary baseline covariate, $Z_1$ is a binary treatment at time 1, $L_1$ is a binary time-varying covariate between time 1 and 2, $Z_2$ is a binary treatment, and $Y$ is a binary outcome. To obtain the posterior distribution of
\begin{eqnarray}
\bE\{ Y(z_1, z_2)\} &=&  \sum_{l_0=0,1} \sum_{l_1=0,1} \Pr(Y=1\mid Z_1=z_1,Z_2=z_2,L_0=l_0, L_1=l_1) \nonumber  \\
&& \quad \quad \quad\quad \quad \quad  \cdot 
\Pr(L_1=l_1\mid Z_1=z_1, L_0=l_0) \cdot  \Pr(L_0=l_0), \label{eq::g-formula-2}
\end{eqnarray}
it suffices to obtain the posterior distributions of the probabilities in \eqref{eq::g-formula-2}. 
Assuming the standard Beta$(1/2,1/2)$ conjugate priors. 
We can obtain the posterior of the probabilities as follows:  
\emph{(i)} sample $\Pr(Y=1\mid Z_1=z_1,Z_2=z_2,L_0=l_0, L_1=l_1)$ from 
Beta$(1/2 + \sum_{i=1}^N \one(Z_{i1}=z_1, Z_{i2}=z_2, L_{i0}=l_0, L_{i1}=l_1)Y_i, 1/2 + \sum_{i=1}^N\one(Z_{i1}=z_1, Z_{i2}=z_2, L_{i0}=l_0, L_{i1}=l_1) (1-Y_i)  )$,
\emph{(ii)} sample $\Pr(L_1=1\mid Z_1=z_1, L_0=l_0)$ from 
Beta$(1/2 + \sum_{i=1}^N \one(Z_{i1} = z_1, L_{i0}=l_0)  L_{i1}, 1/2 + \sum_{i=1}^N \one(Z_{i1} = z_1, L_{i0}=l_0)   (1-L_{i1}) )$, 
and \emph{(iii)} sample $\Pr(L_0=1)$ from Beta$(1/2 + \sum_{i=1}^N L_{i0} , 1/2 + \sum_{i=1}^N (1-L_{i0}) )$.
With these ingredients and \eqref{eq::g-formula-2}, we can obtain the posterior distributions of $\bE\{ Y(z_1, z_2)\} $'s and their contrasts $\sum_{z_1,z_2} c(z_1,z_2) \bE\{ Y(z_1, z_2)\}$. $\Box$

G-computation quickly becomes complex as the number of time periods $T$ and  time-varying confounders increases, which requires specifying a large number of models. Then it is necessary to impose more structural restrictions on the data-generating process. However, Robins and Wasserman \cite{robins1999estimation} showed that unsaturated models might rule out the null hypothesis of zero causal effect {\it a priori}, a phenomenon termed as the ``g-null paradox''.

A popular alternative strategy is the marginal structural model \cite{RHB:2000}, which generalizes IPW to time-varying treatments. Saarela et al. \cite{Saarela15} devised a Bayesian version of the marginal structural model via the Bayesian bootstrap. 
Because the marginal structural model relies on IPW, a key component in its implementation is to estimate the propensity scores and ensure overlap at each time point. However, overlap between different treatment paths usually becomes limited as the number of time periods increases, rendering the marginal structural model to be sensitive to extreme weights.

The above discussion focuses on \emph{static} treatment sequences. Another important class of time-varying treatment is the \emph{dynamic} treatment regime, which consists of a sequence of decision rules, one per time point of intervention, that determines how to individualize treatments to units based on evolving treatment and covariate history. Inferring optimal dynamic treatment regimes requires combining causal inference and decision theory techniques and is closely related to reinforcement learning. See \cite{chakraborty2014dynamic} for a review. 
Due to the space limit, we  omit the discussion of the closely related topics of Bayesian multi-armed bandit \cite{scott2010} and Bayesian reinforcement learning \cite{Ghavamzadeh2015}.

\section{Discussion} \label{sec::discussion}
This paper reviews the Bayesian approach to causal inference under the potential outcomes framework. We discussed the causal estimands, identification strategies, the general structure of Bayesian inference of causal effects, and sensitivity analysis. We highlight issues that are unique to Bayesian causal inference, including the role of the propensity score, definition of identifiability, the choice of priors in both low and high dimensional regimes. In particular, under ignorability and prior independence, the propensity score is seemingly irrelevant for the posterior distributions of the causal parameters. However, we pointed out that even in this setting, the propensity score and more generally the design stage plays a central role in obtaining robust Bayesian causal inference. Regardless of the mode of inference, a critical step in causal inference with observational data is to ensure adequate covariate overlap and balance in the design or analysis stages. In high dimensions, such a task is particularly challenging and what is the optimal practice remains to be an open question.

The Bayesian approach offers several advantages for causal inference. First and most importantly, by enabling imputation of all missing potential outcomes, the Bayesian paradigm provides a unified inferential framework for any causal estimand. This is particularly appealing for inferring complex estimands such as the conditional average treatment effects or individual treatment effects as well as partially identifiable causal estimands such as the principal strata causal effects. In contrast, the Frequentist approach to these problems needs to be customized for each scenario, and the inference usually relies on bounds or asymptotic arguments, which are often either non-informative or questionable in cases like individual treatment effects. Second, the automatic uncertainty quantification of any estimand renders it straightforward to combine causal inference and decision theory for dynamic decision making, e.g. in personalized medicine. Third, the Bayeian approach naturally incorporates prior knowledge into a causal analysis, e.g. in evaluating spatially correlated treatments and/or outcomes. Fourth, there is rich collection of Bayesian models for complex data with limited Frequentists counterparts. A few examples are \emph{(i)} spatial or temporal data, \emph{(ii)} functional data, and \emph{(iii)} interference, i.e. when the SUTVA assumption is violated. In these settings, special care must be taken on issues key to causal inference such as defining relevant estimands and ensuring overlap. Moreover, it is important to ensure that the Bayesian models are coherent to the model-free identification assumptions such as ignorability. For example, adding spatial random effects into an outcome model may inadvertently bias the coefficient of the treatment variable as the estimate of a causal effect \cite{schnell2020mitigating}. Research on Bayesian analysis of these topics has been rapidly increasing \cite{daniels2012bayesian, schnell2020mitigating, zeng2021causal,forastiere2022estimating, linero2022how} and is expected to continue to grow. 

Despite the above advantages, the theory and practice in causal inference has long been dominated by non-Bayesian methods. One reason is that many popular Frequentist techniques, such as matching and weighting, as well as Fisherian randomization test, do not require specifying outcome models and prior distribution of parameters, and thus offer a perception of ``model-free'' or ``objective''. This is appealing to many applied researchers. Another reason is that the Bayesian approach requires more advanced computing and programming, which may not be readily available to many practitioners. The \textbf{Stan} programming language \cite{stan} mitigates some of these issues, but Bayesian computation remains to be inaccessible to most domain scientists. To popularize Bayesian causal inference in practice, it is crucial to provide \emph{(i)} more examples of successful Bayesian applications with clear advantages over other inferential modes, e.g. \cite{dorie2019automated}, \emph{(ii)} accessible tutorials, ideally with generalizable computer code and illustrations of important scientific problems, and \emph{(iii)} developing and disseminating user-friendly, general purpose software packages.

We have occasionally commented whether a method is dogmatically Bayesian in the discussion. However, we do not regard the conceptual purity of being dogmatically Bayesian, \emph{per se}, is advantageous, nor should it be the motivating goal in real applications. When a quasi-Bayesian method outperforms its dogmatically Bayesian counterpart (if available) with methodological footing and empirical evidence, as the example of adding estimated propensity score in an outcome model in Section \ref{sec::PSrole}\ref{sec::PSrole-pscovariate}, we would endorse the former over the latter. We also doubt the value of devising a Bayesian version of an established Frequentist method without clear theoretical or practical advantages. As a general view, we believe whether to choose a Bayesian approach should be dictated by its practical utility in a specific context rather than an unconditional commitment to the Bayesian doctrine. For causal inference and perhaps everything in statistics, being Bayesian should be a tool, not a goal.

\vskip6pt

\enlargethispage{20pt}


\dataccess{This study does not involve data.}

\aucontribute{FL conceived of and designed the study, and FL, PD, FM drafted the manuscript. All authors read and approved the manuscript.}

\competing{The authors declare that they have no competing interests.}

\funding{PD's research is partially funded by the U.S. National Science Foundation grant \# 1945136. FM thanks the Department of Excellence 2018-2022 funding provided by the
Italian Ministry of University and Research (MUR).}

\ack{The authors are grateful to Joey Antonelli, Yuansi Chen, Sid Chib,  Ruobin Gong, Guido Imbens, Zhichao Jiang, Antonio Linero, Georgia Papadogeorgou, Donald Rubin,  Surya Tokdar, Mike West, Jason Xu,  Cory Zigler, Anqi Zhao, and two anonymous reviewers for discussions and suggestions.}

\disclaimer{None.}



\end{document}